\documentclass[conference]{IEEEtran}
\IEEEoverridecommandlockouts

\usepackage{cite}
\usepackage{amsmath}
\usepackage{algpseudocode}
\usepackage[ruled,vlined,linesnumbered]{algorithm2e}
\usepackage{amsmath,amssymb,amsfonts}
\usepackage{graphicx}
\usepackage{textcomp}
\usepackage{xcolor}
\usepackage{svg}
\usepackage{subcaption}
\usepackage{mwe}
\usepackage[a4paper,
            left=0.75in,
            right=0.75in,
            top=0.7in,
            bottom=1.6in]{geometry}

\usepackage[font=small]{caption}

\let\oldnl\nl
\newcommand{\nonl}{\renewcommand{\nl}{\let\nl\oldnl}}

\def\BibTeX{{\rm B\kern-.05em{\sc i\kern-.025em b}\kern-.08em
    T\kern-.1667em\lower.7ex\hbox{E}\kern-.125emX}}
\begin{document}

\title{\vspace{-1mm} Foundation Model-Aided Deep Reinforcement Learning for RIS-Assisted Wireless Communication\\ \vspace{-1mm}}

\author{\IEEEauthorblockN{Mohammad Ghassemi\textsuperscript{1}, Sara Farrag Mobarak\textsuperscript{1}, Han Zhang\textsuperscript{1}, Ali Afana\textsuperscript{2}, Akram Bin Sediq\textsuperscript{2}, \\ and Melike Erol-Kantarci\textsuperscript{1}, \textit{Fellow}, \textit{IEEE}}
\IEEEauthorblockA{\textit{\textsuperscript{1}School of Electrical Engineering and Computer Science, University of Ottawa, Ottawa, Canada} \\
\textit{\textsuperscript{2}Ericsson Inc., Ottawa, Canada}\\
Emails:\{mghas017, smobarak, hzhan363, melike.erolkantarci\}@uottawa.ca,
\{ali.afana, akram.bin.sediq\}@ericsson.com}

\vspace{-8mm}}

\maketitle

\begin{abstract}
Reconfigurable intelligent surfaces (RIS) have emerged as a promising technology for enhancing wireless communication by dynamically controlling signal propagation in the environment. However, their efficient deployment relies on accurate channel state information (CSI), which leads to high channel estimation overhead due to their passive nature and the large number of reflective elements. In this work, we solve this challenge by proposing a novel framework that leverages a pre-trained open-source foundation model (FM) named large wireless model (LWM) to process wireless channels and generate versatile and contextualized channel embeddings. These embeddings are then used for the joint optimization of the BS beamforming and RIS configurations. To be more specific, for joint optimization, we design a deep reinforcement learning (DRL) model to automatically select the BS beamforming vector and RIS phase-shift matrix, aiming to maximize the spectral efficiency (SE).
This work shows that a pre-trained FM for radio signal understanding can be fine-tuned and integrated with DRL for effective decision-making in wireless networks. It highlights the potential of modality-specific FMs in real-world network optimization.
According to the simulation results, the proposed method outperforms the DRL-based approach and beam sweeping-based approach, achieving 9.89\% and 43.66\% higher SE, respectively.
\end{abstract}



\section{Introduction}
Reconfigurable intelligent surfaces (RISs) are a groundbreaking technology for future wireless networks that dynamically control and optimize electromagnetic wave propagation to optimize signal propagation \cite{zhou2024cooperative}. Channel estimation serves as a fundamental enabler in RIS-assisted systems, providing the necessary CSI for optimizing phase-shift configurations and enhancing spectral efficiency (SE). 
However, it poses significant challenges due to the passive nature of RIS elements and the large number of reflecting elements. This makes direct estimation of channels between the base station (BS) and RIS, as well as channel estimation between RIS and users, challenging \cite{zhou2022channel} \cite{alikhani2024digital}. Existing RIS channel estimation methods often suffer from high pilot training overhead, significant power leakage, large estimation errors, or reliance on impractical assumptions \cite{mobarak2025sum}. 
To address these challenges, prior work has explored structured sparsity and compressive sensing for RIS-assisted mmWave systems to reduce pilot overhead and improve scalability \cite{zhou2022channel}. Additionally, beam-sweeping techniques have been studied to identify optimal beams without requiring explicit channel state information (CSI). For instance, a hierarchical codebook-based beam training framework was proposed to enhance beam alignment efficiency and reduce training overhead in RIS-assisted mmWave communication \cite{wang2023hierarchical}. However, in multi-user RIS-aided networks, beam sweeping often results in extensive over-the-air computations and interactions with users and needs very large codebooks for interference management.

Deep reinforcement learning (DRL) has emerged as a powerful tool for optimizing complex decision-making processes in wireless communication systems. In wireless communication contexts, DRL overcomes obstacles by utilizing its capacity to learn optimal policies by interaction with dynamic environments. For instance, the authors in \cite{zhu2022drl} proposed a DRL-based framework for joint beamforming and RIS phase shift optimization in mmWave networks, achieving significant performance gains. Similarly, \cite{abdallah2024multi} introduced a multi-agent DRL approach for dynamic RIS-assisted networks, enabling real-time adaptation to changing channel conditions. Despite these advancements, the effectiveness of DRL in RIS-assisted networks heavily depends on the accuracy of CSI. Inaccurate CSI can mislead the DRL agent, resulting in suboptimal policy learning and degraded network performance. 

The Large Wireless Model (LWM) is a foundation model (FM) developed for wireless communication and sensing \cite{touvron2023llama} \cite{alikhani2024large}.
It incorporates a self-attention mechanism, enabling it to capture long-range dependencies in the wireless environment to produce rich and contextualized channel embeddings. This capability enhances its performance across diverse wireless communication and sensing tasks \cite{alrabeiah2020deep} \cite{ghassemi2025generative}.
As a result, it can help with our RIS-assisted communication task by providing more generalizable channel representations, ultimately leading to SE gains.

Building on this foundation, we propose the FM-aided DRL (FMDRL) method that combines LWM with DRL to further enhance the adaptability and efficiency of RIS-assisted wireless systems. The key contributions of this work are summarized as follows:

\begin{itemize}
\item We fine-tune the LWM \cite{alikhani2024large} for RIS-assisted scenarios, adapting it to provide accurate and context-aware CSI tailored to the task of channel estimation.
\item We propose a novel framework that integrates the fine-tuned LWM with DRL to jointly optimize BS beamforming and RIS phase element configurations for real-world decision-making tasks.
\item Our approach improves the SE, demonstrating the practical benefits of combining FMs with DRL in complex wireless environments.
\end{itemize}

The structure of the paper is organized as follows.
Section II presents the related work. Section III outlines the system model and problem formulation. Section IV presents our approach to solve the problem. Section V presents simulation results and performance comparisons. Finally, Section VI concludes the study by summarizing the findings and implications of the proposed approach.


\section{Related work}

Several existing works have addressed optimization in RIS-assisted networks using both traditional and learning-based approaches. For example, \cite{jin2022ris} formulated a downlink system design that jointly optimizes BS beamforming and RIS phase shifts to minimize transmit power using iterative optimization techniques. DRL has also been investigated as a tool for dynamic control in RIS-assisted systems. In \cite{jiao2021enabling}, RIS was applied to mitigate dynamic blockages in high-frequency networks, showcasing its adaptability to varying environments. In \cite{tseng2025joint}, the authors employed DRL to jointly optimize beamforming and resource allocation, while \cite{lee2020deep} proposed a DRL-based framework for energy-efficient transmission by dynamically adjusting RIS and BS parameters based on real-time CSI and energy availability.

Recent works have also explored the use of FMs, such as the LWM \cite{alikhani2024large}, to generate universal, context-aware channel embeddings that improve performance across various downstream tasks. These methods leverage the representational power of pre-trained models to replace or augment traditional CSI, improving task performance. Similarly, the Wireless Foundation Model (WiFo) \cite{liu2024wifo} employs a masked autoencoder architecture to handle heterogeneous space-time-frequency CSI data, enabling zero-shot generalization across various channel prediction tasks without fine-tuning.
However, these works typically focus on supervised or regression-based applications. Their integration into optimization frameworks remains unexplored. In contrast, our approach integrates LWM with DRL, using the rich, pre-trained channel embeddings as input to guide decision-making in a DRL agent. This combination of FMs with DRL facilitates adaptive, real-time control in RIS-assisted networks, extending the applicability of FMs beyond passive inference to active decision-making processes.


\section{System Model and Problem Formulation}
\subsection{System Model}
\begin{figure}
\centering
  \centering
  \includegraphics[width=1.0\linewidth]{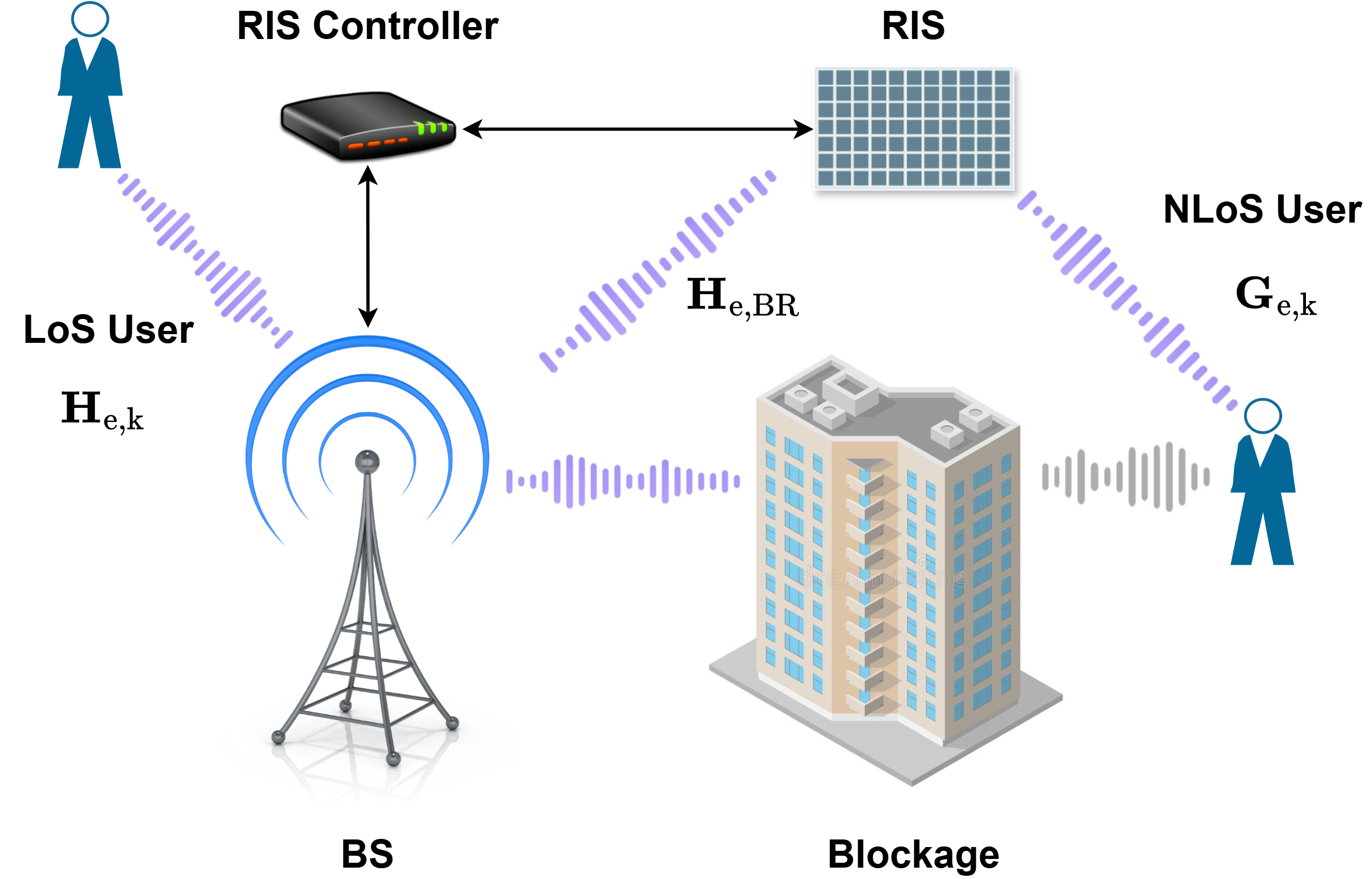}
  \captionof{figure}{System architecture for FMDRL.}
  \label{fig:image1}
\end{figure}

In this work, a RIS-aided wireless communication system is considered, as illustrated in \figurename{\ref{fig:image1}}. 
The proposed system consists of a BS equipped with $N_t$ antennas, a RIS comprising $M$ reflecting elements, and $K$ User Equipment (UE) devices, each with $N_r$ receive antennas.
The BS serves UEs via line-of-sight (LoS) transmissions, while the RIS creates enhanced propagation paths by reconstituting non-line-of-sight (NLoS) components into quasi-LOS links. With this setup, we aim to enhance the SE with the proposed FMDRL framework.

\subsection{RIS-Aided Wireless Communication Model}

In the following, we explain how we incorporate the embedded channel representations into both the direct transmission and the RIS-assisted reflection. This model also considers the multi-user interference and RIS phase shifts, which are essential for accurate signal reconstruction in complex wireless environments. The received signal $\mathbf{y}_k$ at the $k$th user is given by:
\begin{equation} 
\mathbf{y}_k = (\mathbf{H}_{e,k} + \mathbf{G}_{e,k} \mathbf{\Theta} \mathbf{H}_{e,\text{BR}}) \sum_{u=1}^{K} \mathbf{f}_u x_u + \mathbf{n}_k,
\end{equation}  
where $\mathbf{H}_{e,k} \in \mathbb{C}^{N_r \times N_t}$ is the embedded direct channel between the BS and the $k$th user. The subscript $e$ denotes an embedding transformation applied to the original channel representation. $\mathbf{H}_{e,\text{BR}} \in \mathbb{C}^{M \times N_t}$ represents the channel from the BS to the RIS, and $\mathbf{G}_{e,k} \in \mathbb{C}^{N_r \times M}$ is the channel from the RIS to the $k$th user. The term $x_u$ refers to the transmitted symbol for user $u$, and $\mathbf{n}_k$ is the additive noise at the $k$th user.
The RIS is used to improve communication by dynamically adjusting its phase shifts to enhance the signal quality between the BS and users. These phase shifts are modeled by a diagonal phase-shift matrix $\mathbf{\Theta} = \text{diag}(\boldsymbol{\theta})$, containing the RIS phase shifts on its diagonal, where:
\begin{equation} 
\boldsymbol{\theta} = [e^{j\phi_1}, e^{j\phi_2}, \dots, e^{j\phi_M}],
\end{equation}  
represents the adjustable phase shifts with $|\theta_m| = 1$ for $m = 1, \dots, M$.
In the received signal expression, $\mathbf{f}_u \in \mathbb{C}^{N_t \times 1}$ is the beamforming vector for the $u$th user, where the full beamforming matrix is denoted by $\mathbf{F} = [\mathbf{f}_1, \mathbf{f}_2, \dots, \mathbf{f}_K] \in \mathbb{C}^{N_t \times K}$, and $\mathbf{n}_k \sim \mathcal{CN}(0, \sigma_k^2 \mathbf{I})$ represents the additive white Gaussian noise (AWGN) at user $k$.  
The SE for the $k$th user is given by:
\begin{equation} 
\resizebox{.89\hsize}{!}{$\text{SE}_k = \log_2 \left( 1 + \frac{\left\| (\mathbf{H}_{e,k} + \mathbf{G}_{e,k} \mathbf{\Theta} \mathbf{H}_{e,\text{BR}}) \mathbf{f}_k \right\|^2}{\sum\limits_{u \neq k} \left\| (\mathbf{H}_{e,k} + \mathbf{G}_{e,k} \mathbf{\Theta} \mathbf{H}_{e,\text{BR}}) \mathbf{f}_u \right\|^2 + \sigma_k^2} \right)$}.
\end{equation}

The objective is to jointly optimize the BS beamforming matrix $\mathbf{F}$ and the RIS phase-shift matrix $\mathbf{\Theta}$ in order to maximize the system SE. This joint design problem can be formulated as the following optimization:

\begin{equation} \label{eq:6}
\begin{aligned}
&\max_{\mathbf{F}, \mathbf{\Theta}} \sum_{t=1}^{T} \left( \sum_{k=1}^{K} \text{SE}_k\right)^t \\ 
& \,\,\,\, \text{s.t.} \quad \left\|\mathbf{F}\right\|^2 \leq P_{max},\\
& \,\quad\quad\quad |\theta_m| = 1, \,\, \theta_m \in [0, 2\pi), \; \,\, m \in M,
\end{aligned}
\end{equation}
where $T$ denotes the total number of time steps per episode in DDPG-based training, and $P_{max}$ represents the maximum power of the BS.
This formulation jointly optimizes the BS beamforming matrix $\mathbf{F}$ and the RIS phase-shift matrix $\mathbf{\Theta}$ to maximize the total system SE while satisfying the constraints.

\section{FM-Aided DRL for RIS-Assisted Wireless Communication}
\begin{figure*}
\centering
  \centering
  \includegraphics[width=1\linewidth]{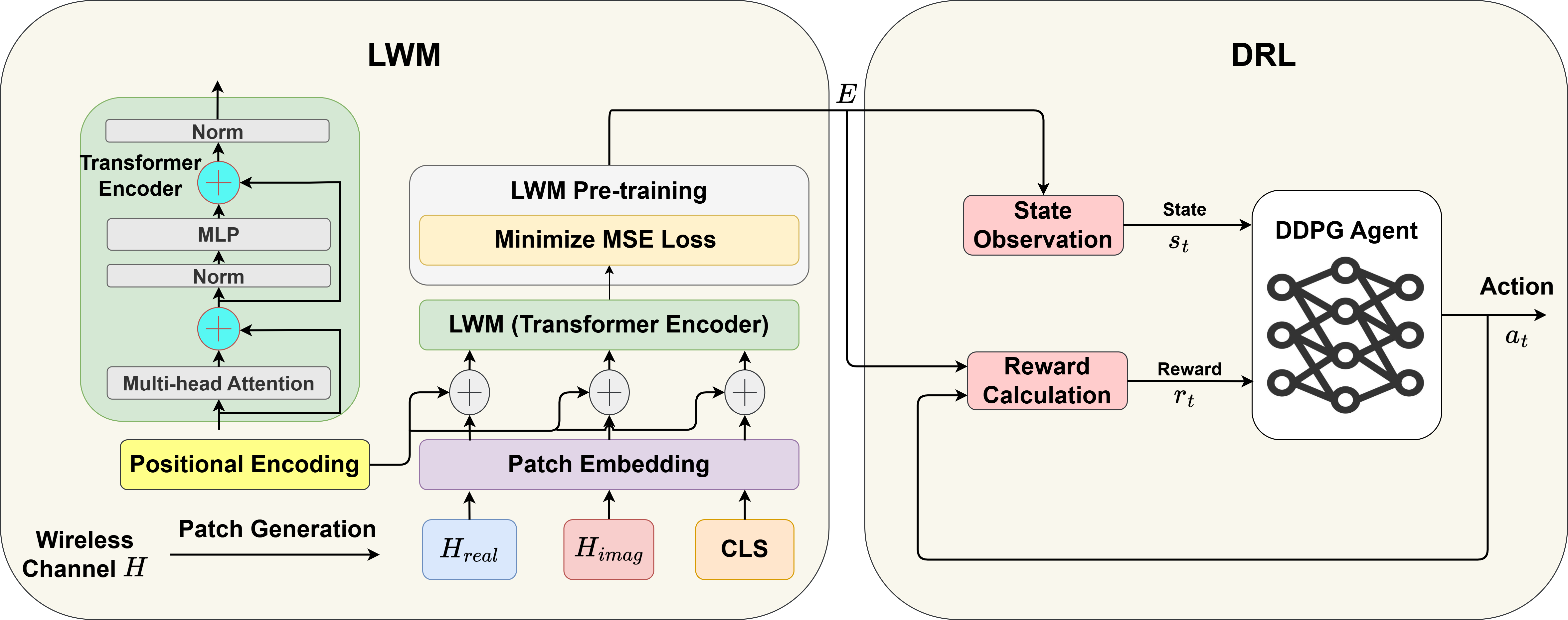}
  \captionof{figure}{System architecture for optimizing the BS beamforming matrix and the RIS phase-shift matrix by leveraging FMDRL}
  \label{fig:image2}
\end{figure*}

In this section, we present the detailed implementation of our proposed FMDRL framework, which combines the LWM with DRL agent, as illustrated in \figurename{\ref{fig:image2}}. The LWM extracts rich, context-aware embeddings from the wireless channel, which are then used by the DRL agent to make optimal control decisions based on the observed state and received rewards.
Since the objective function and the constraints in \ref{eq:6} are non-convex, leading to a non–convex non–trivial optimization problem.
In this work, instead of tackling the complex optimization problem through mathematical derivation, we reformulate the SE optimization as a deep deterministic policy gradient (DDPG)-based DRL problem to derive feasible solutions for the BS beamforming matrix $\mathbf{F}$ and the RIS phase-shift matrix $\mathbf{\Theta}$.

\subsection{Large Wireless Model (LWM)}
The proposed framework leverages the LWM, a transformer-based FM  pre-trained on large-scale wireless channel datasets using self-supervised learning. LWM takes raw wireless channel matrices as input and produces rich, contextualized channel embeddings as output. These embeddings capture both spatial and spectral features of the wireless environment, enabling improved performance in downstream tasks such as channel estimation, beamforming, and interference management \cite{ghassemi2024multi} \cite{farzanullah2025beam}.

To adapt the pre-trained LWM to RIS-assisted wireless communication tasks, we fine-tune it using masked self-supervised learning. 
To prepare the wireless channel for transformer-based processing, the raw channel matrix is first separated into its real and imaginary parts. Given a channel matrix $\mathbf{H} \in \mathbb{C}^{X \times Y}$, the real and imaginary components are extracted as $\mathbf{H}_{\text{real}} = \Re(\mathbf{H}), \mathbf{H}_{\text{imag}} = \Im(\mathbf{H})$.
We then flatten each matrix by applying the vectorization operator to their transposes $\mathbf{h}_{\text{real}} = \text{vec}(\mathbf{H}_{\text{real}}^\top), \mathbf{h}_{\text{imag}} = \text{vec}(\mathbf{H}_{\text{imag}}^\top)$.
This representation allows the channel to be divided into fixed-size patches, forming the input tokens for the transformer encoder.
The input is divided into $P$ patches in total, with $P/2$ patches for the real component and $P/2$ for the imaginary component. Each patch has a fixed length of $L = \frac{2XY}{P}$ as:
\begin{equation} 
\begin{aligned}
& \mathbf{p}_i = \mathbf{h}_{\text{real}}[(i-1)L : iL],  \quad i \in \left\{1, \dots, \frac{P}{2} \right\} \\  & \mathbf{p}_{i + P/2} = \mathbf{h}_{\text{imag}}[(i-1)L : iL],   \quad i \in \left\{1, \dots, \frac{P}{2} \right\},
\end{aligned}
\end{equation}
where each patch $\mathbf{p}_i \in \mathbb{R}^L$. Also, the classification (CLS) token is a special token added to the input sequence, whose final hidden state is often used as a fixed-length representation of the entire sequence for classification tasks. This CLS token aggregates information across all patches and provides a global representation of the channel. The CLS patch increases the sequence length to $ P + 1 $. 
 
During fine-tuning, we apply a masked channel modeling strategy similar to the one used in pre-training. Specifically, 15\% of the input patches are randomly selected for masking. Among these, 80\% are replaced with a mask token, 10\% with random noise, and 10\% remain unchanged. The model is trained to reconstruct the masked patches using the Mean Squared Error (MSE) loss:
\begin{equation} 
\mathcal{L} = \frac{1}{|\mathbf{M}_p|} \sum_{i \in \mathbf{M}_p} \left\| W_{\text{dec}} e_i^{\text{LWM}} - \mathbf{p}_i \right\|^2,
\end{equation}
where $\mathbf{M}_p$ is the set of masked patches, $ \mathbf{W}_{\text{dec}} $ is a learnable projection matrix, $ \mathbf{e}_i^{\text{LWM}} $ is the LWM embedding of the masked patch, and $\mathbf{p}_i$ is the original patch. This fine-tuning process enables the LWM to capture RIS-specific patterns, improving its ability to provide accurate and generalizable channel embeddings for downstream tasks.

\subsection{Proposed MDP}
To set up the DRL model, we first model the joint optimization of the RIS phase shifts and BS beamforming as a Markov Decision Process (MDP), where the agent interacts with the environment to learn an optimal policy. The objective is to maximize the long-term performance of the system by dynamically adjusting the BS beamforming vectors and RIS configurations based on the observed channel states. The key components of the MDP are defined as follows:
\begin{itemize}
\item \textbf{State Space:} The state in the MDP captures the essential information needed for decision-making, including channel conditions and system configurations. Specifically, the state at time step $t$ is defined as $s_t = \{ \mathbf{H}_{e,k}, \mathbf{G}_{e,k}, \mathbf{H}_{e,\text{BR}} \}$, where $\mathbf{H}_{e,k}$ denotes the direct channel between the BS and user $k$, $\mathbf{G}_{e,k}$ represents the RIS-to-user channel, and $\mathbf{H}_{e,\text{BR}}$ is the BS-to-RIS channel.
\item \textbf{Action Space:} The action involves adjusting the BS beamforming matrix and the RIS phase-shift vector, which can be denoted as $a_t = \{\mathbf{F}, \boldsymbol{\Theta} \}$. These parameters directly influence the $\text{SE}_k$.
\item \textbf{Reward:} The reward is designed to encourage the agent to maximize the SE, which can be denoted as $r_t = \sum_{t=1}^{T} \text{SE}_t$.
\end{itemize}

We adopt the DDPG algorithm to solve this MDP. DDPG is an actor-critic reinforcement learning algorithm that operates in continuous action spaces, making it well-suited for optimizing the BS beamforming matrix $\mathbf{F}$ and the RIS phase-shift matrix $\mathbf{\Theta}$. 
The actor network generates actions based on the observed state, whereas the critic network assesses the quality of those actions by estimating their Q-values.
Through experience replay and target network updates, DDPG efficiently learns an optimal policy that maximizes the SE in a dynamic wireless environment.

\subsection{Training Process of DDPG}
The training of the DDPG agent aims to learn a policy that jointly optimizes the BS beamforming matrix $\mathbf{F}$ and the RIS phase-shift matrix $\boldsymbol{\Theta}$. It begins by initializing the environment with an initial state $s_0 = E$, which is generated from the output of the LWM module. At each time step $t$, the agent observes the current state $s^{(t)}$ and selects a continuous action $a^{(t)}$ using the actor network. This action, which includes both $\mathbf{F}$ and $\boldsymbol{\Theta}$, is scaled to meet the power constraints defined in (\ref{eq:6}). The agent then executes the action, receives a reward $r_t$ based on the resulting SE, and observes the next state $s^{(t+1)}$, which is also normalized. The transition tuple $(s^{(t)}, r_t, s^{(t+1)})$ is stored in a replay buffer \cite{mobarak2025sum}.
Throughout the training, mini-batches are sampled from the replay buffer to update the actor and critic networks via gradient-based optimization. Penalty terms are introduced when constraints are violated, encouraging the agent to learn feasible and efficient solutions. This iterative process continues until the policy converges to optimal configurations, yielding the final beamforming and phase-shift matrices, $\mathbf{F}_{\text{opt}}$ and $\boldsymbol{\Theta}_{\text{opt}}$.
The complete training procedure is summarized in Algorithm 1.

\begin{algorithm}[htbp]
\caption{Training Procedure for Joint BS Beamforming Matrix $\mathbf{F}$ and RIS Phase-Shift Matrix $\boldsymbol{\Theta}$ Using DDPG}
\SetAlgoNlRelativeSize{-1}
\KwIn{Initial state $s_0 = E$, actor and critic networks, and network parameters (e.g., $P_{\max}$, BS beamforming $\mathbf{F}$, RIS phase-shift matrix, etc.)}
\KwOut{BS beamforming matrix $\mathbf{F}_{\text{opt}}$ and RIS phase-shift matrix $\boldsymbol{\Theta}_{\text{opt}}$}

\For{$t = 1$ to $T$}{
    Reset environment: $s^{(t)} \leftarrow \text{normalize}(s^{(t)})$\;
    Select action $a^{(t)}$ using actor network\;
    $a^{(t)} \leftarrow \text{scale\_actions}(a^{(t)})$\;
    Execute action $a_t$, observe reward $r_t$\; 
    Compute next state $s^{(t+1)}$\;
    $s^{(t+1)} \leftarrow \text{normalize}(s^{(t+1)})$\;
    Store transition $(s^{(t)}, r_t, s^{(t+1)})$\;
    Sample mini-batch\;
    Update network\;
    Apply penalties if constraints are violated\;
    $s^{(t)} \leftarrow s^{(t+1)}$\;
}
\nonl
$\mathbf{Return}$ BS beamforming matrix $\mathbf{F}_{\text{opt}}$ and RIS phase-shift matrix $\mathbf{\Theta}_{\text{opt}}$\;
\label{alg:FMDRL}
\end{algorithm}


\section{Simulation and Results}
\subsection{Parameter Settings}

The LWM used in this work follows the original architecture proposed in \cite{alikhani2024large}.
We fine-tune the last layer of the LWM model on a realistic outdoor urban scenario from the DeepMIMO dataset \cite{alikhani2024digital},\cite{alkhateeb2019deepmimo} with batch size = 64 and learning rate = $1 \times 10^{-5}$, employing MSE loss on complex-valued channel coefficients.  The model operates with a patch size of $L = $ 32 and channel patches of $P = $ 32. 
it uses AdamW optimization with learning rate = $5 \times 10^{-5}$, $\beta_1$ = 0.9, $\beta_2$ = 0.999, $\epsilon = 1 \times 10^{-8}$, weight decay of $1 \times 10^{-5}$, and layer normalization for stable training, with a batch size of 64.

The simulation setup consists of $K$ users with the number of receive antennas $N_r = $ 1, a RIS with $M = $ 32 elements, a BS with $N_t = $ 32 antennas, and a power of $P_{max}$. 
The DeepMIMO dataset generator \cite{alkhateeb2019deepmimo} is configured based on the specified parameters and system settings, simulating a realistic wireless channel with multiple propagation paths. The system operates with a bandwidth of 0.5 GHz as a frequency-flat system.

For DDPG, the learning rates for the actor and critic networks are set to $10^{-4}$ and $10^{-3}$, respectively, with a discount factor of 0.99 and a batch size of 64. The training is conducted over $T = 150,000$ episodes. The target networks are updated with a soft update rate of 0.005. The neural networks have two hidden layers with dimensions $\{64,64\}$. Each episode consists of 100 time steps. The noise variance is $-90$ dBm. 

\subsection{Outdoor Urban Scenario}
We utilize the realistic outdoor (O1) urban scenario from the DeepMIMO dataset to fine-tune the LWM for a wireless communication setting. The environment contains over one million candidate user locations, covering both LOS and NLOS conditions.
We allocate 70\% of the data for model fine-tuning, 15\% for validation, and the remaining 15\% for inference and evaluation. This complex and realistic urban environment makes it a suitable dataset for adapting the FM to RIS-assisted scenarios and benchmarking its performance in practical deployment conditions.

\subsection{Simulation Results}
\begin{figure}[t]
    \centering
    \includegraphics[width=1.0\linewidth]{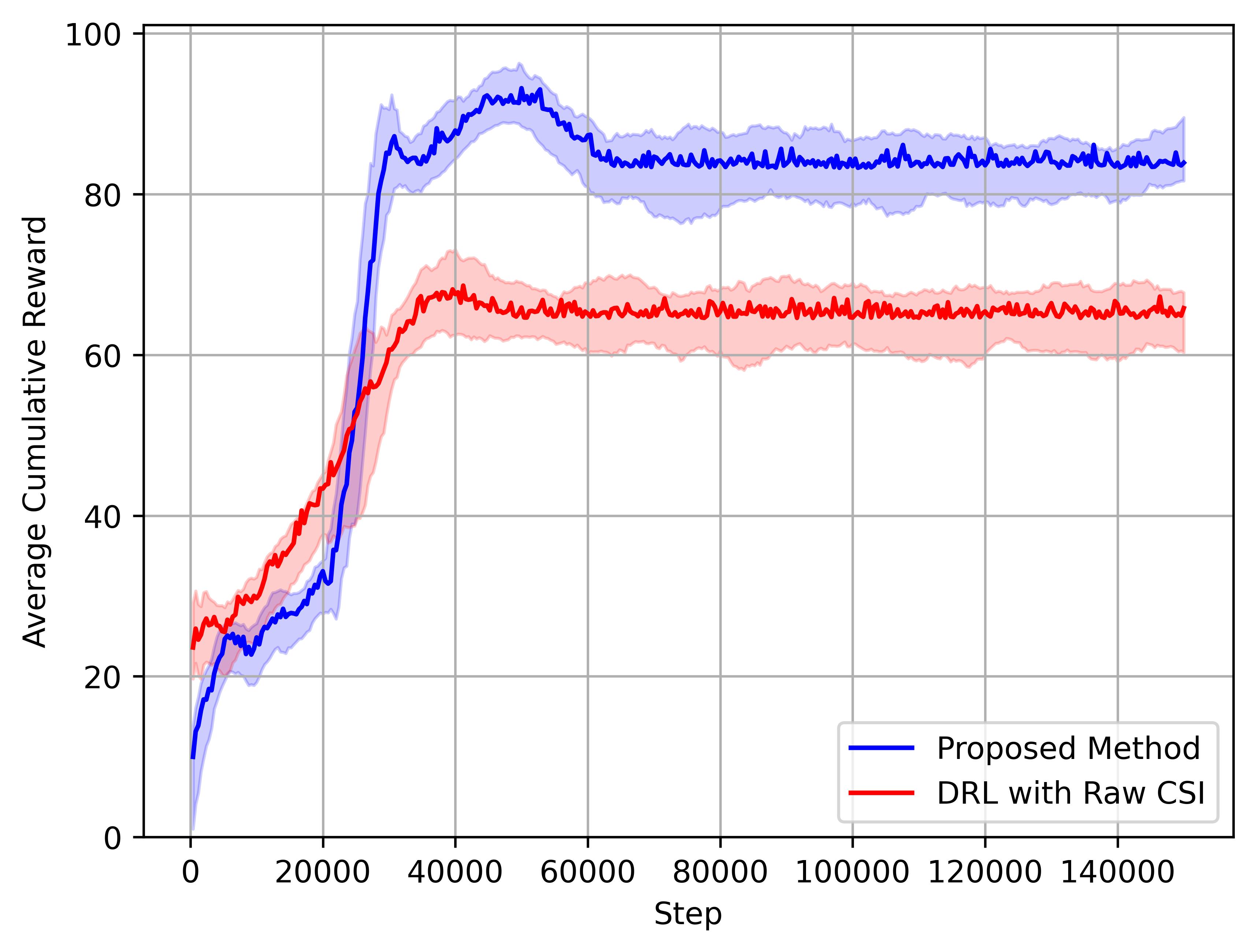}
    \caption{Comparison of average cumulative reward between the FMDRL and DRL with raw CSI.}
    \label{fig:image3}
\end{figure}

In this subsection, we illustrate the performance of the proposed FMDRL approach in jointly optimizing the BS beamforming and phase shift configuration to maximize the SE. There are $K = $ 10 users, and the BS transmit power is  $P_{max} = $  35 dBm.
The results, illustrated in \figurename{\ref{fig:image3}}, show the average cumulative reward over training steps. The FMDRL significantly outperforms the DRL with raw CSI, achieving a higher cumulative reward and demonstrating better convergence behavior. Here, raw CSI refers to unprocessed channel matrices without any learned embedding or contextual feature extraction, as defined in \cite{alikhani2024large}. This improvement highlights the effectiveness of our approach in optimizing beamforming and RIS phase shift configuration for enhanced wireless communication performance.

\begin{figure}[h]
    \centering
    \includegraphics[width=1.0\linewidth]{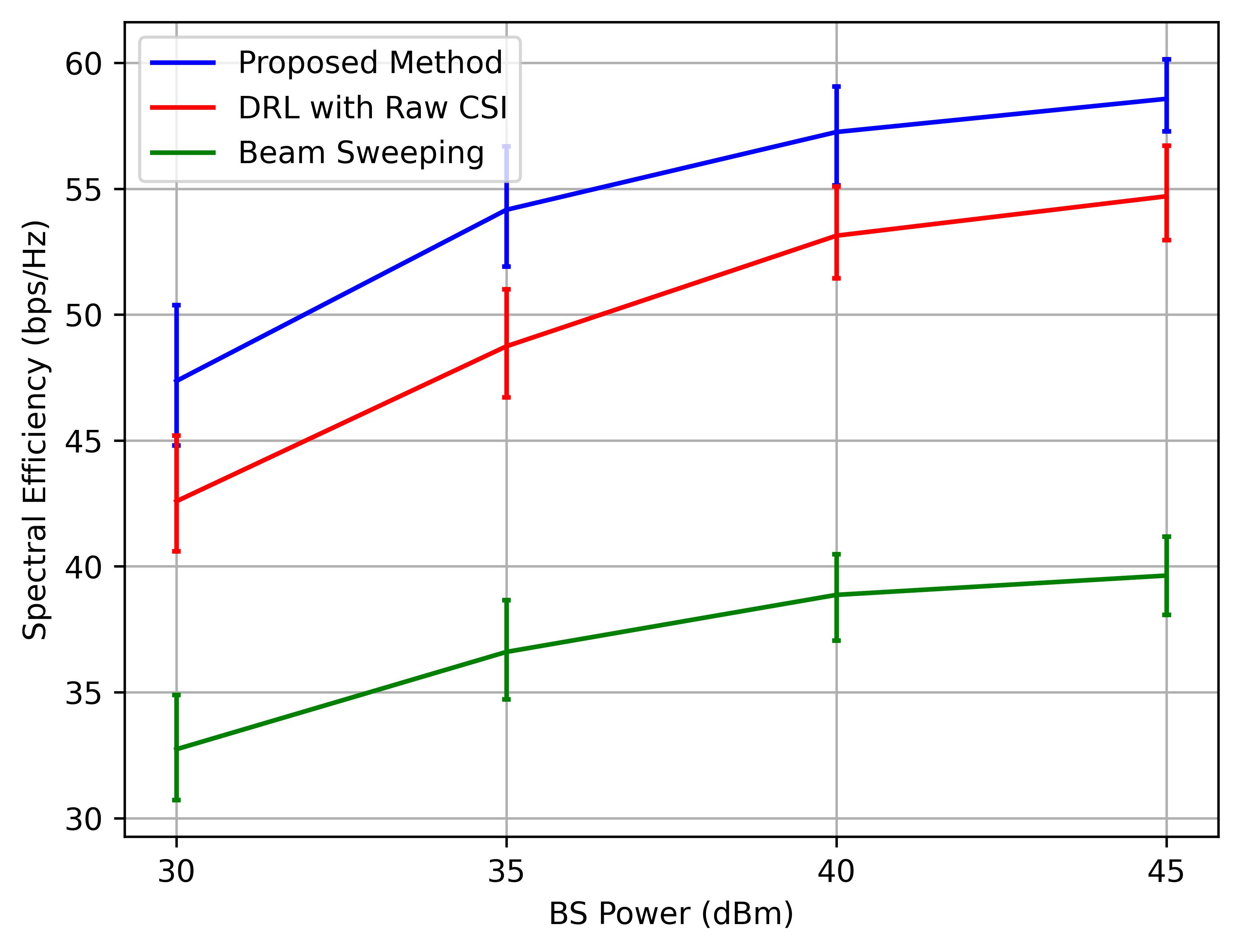}
    \caption{SE performance comparison of the FMDRL, DRL with raw CSI, and beam sweeping for different BS Power.}
    \label{fig:image4}
\end{figure}

We evaluate our proposed RIS-aided system in a scenario with $K = $ 10 users and the BS transmit power $P_{max}$ varied across 30, 35, 40, and 45 dBm to assess performance scalability. For beam sweeping method, we implement beam sweeping with 32 BS beamforming codebooks and 32 RIS codebooks, resulting in 1,024 beam pairs to explore the optimal configuration.
\figurename{\ref{fig:image4}} illustrates the SE performance across different BS power levels for three methods: the proposed approach, DRL with raw CSI, and beam sweeping. 
At 30 dBm, the FMDRL achieves 48.28 bps/Hz, which represents a 9.89\% improvement over DRL with raw CSI with 43.93 bps/Hz and a 43.66\% improvement over beam sweeping with 33.63 bps/Hz. 
When the BS power is increased to 45 dBm, the SE values rise to 58.92 bps/Hz for the FMDRL, showing a 7.16\% improvement over DRL with raw CSI with 54.98 bps/Hz and a 50.02\% increase compared to beam sweeping with 39.27 bps/Hz.
The results demonstrate that the proposed approach consistently outperforms the baseline methods, achieving higher SE across all power levels.

\begin{figure}[t]
    \centering
    \includegraphics[width=1.0\linewidth]{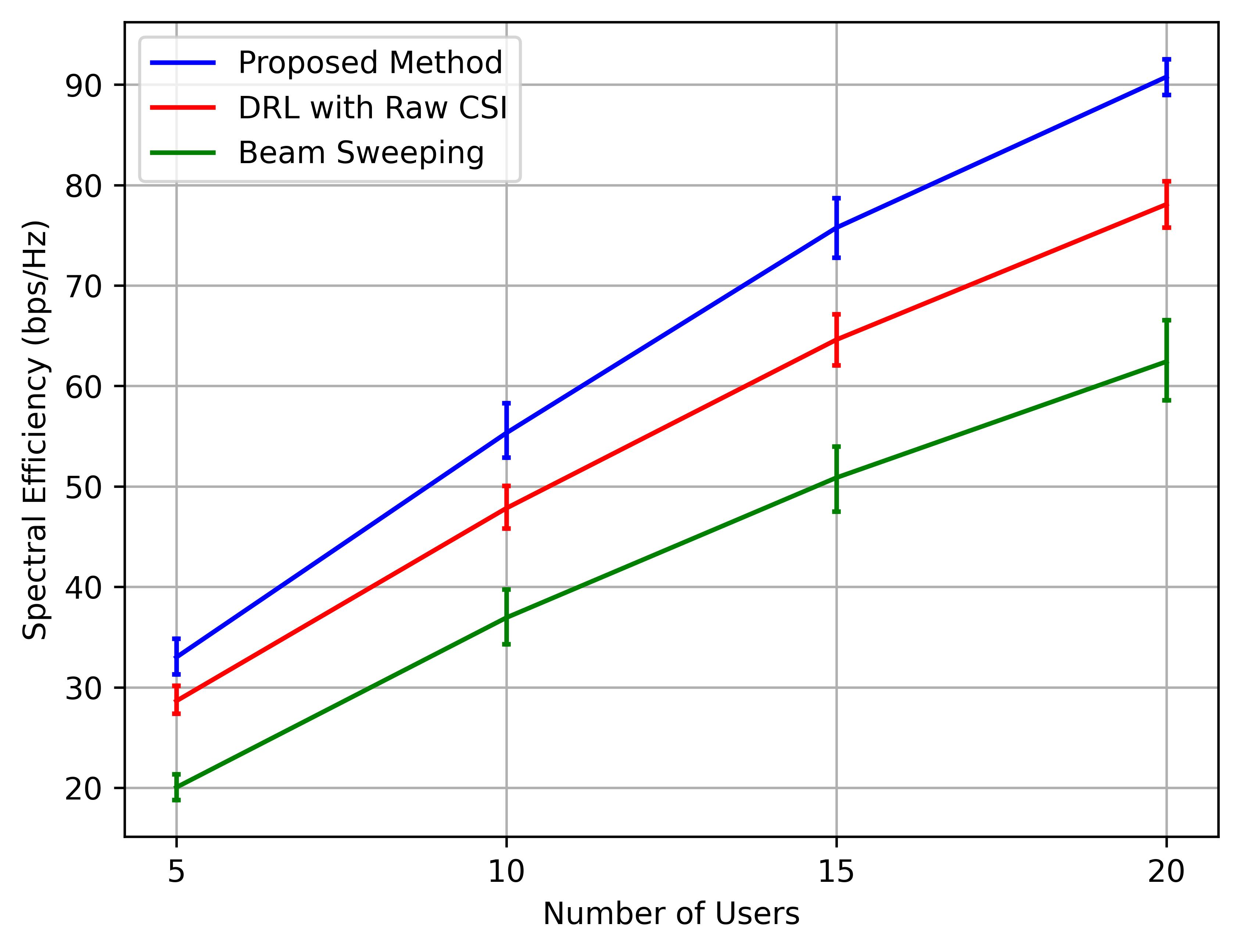}
    \caption{SE performance comparison of the FMDRL, DRL with raw CSI, and beam sweeping for different numbers of users. }
    \label{fig:image5}
\end{figure}

\figurename{\ref{fig:image5}} illustrates the SE performance across different numbers of users to study scalability. We vary the number of users $K$ to 5, 10, 15, and 20. 
The simulation is configured with a BS power of $P_{max}$ =  35 dBm. The beam sweeping setting remains as before. 
At 5 users, the SE values are 33.74 bps/Hz for the FMDRL, achieving a 15.66\% improvement over DRL with raw CSI with 29.17 bps/Hz, and 62.71\% improvement over beam sweeping with 20.74 bps/Hz. 
When the number of users increases to 20, the SE values rise to 91.06 bps/Hz for the FMDRL, representing a 15.48\% improvement over DRL with raw CSI with 78.84 bps/Hz, and 45.52\% increase compared to beam sweeping with 62.59 bps/Hz.
As the number of users increases, the FMDRL maintains a significant performance gain over the baseline approaches, highlighting its effectiveness in optimizing SE.

\section{Conclusion}
In this paper, we propose FMDRL, a DRL-based beam management framework for RIS-assisted wireless networks that combines a fine-tuned LWM for efficient CSI estimation. 
The optimization problem for jointly designing BS beamforming and RIS phase elements is formulated to maximize communication SE.
Experimental results across varying user densities and transmit powers showed consistent superiority over both DRL with raw CSI and exhaustive beam sweeping. The proposed solution addresses critical challenges in RIS deployment, offering a practical pathway for real-world implementation in future wireless networks.

\section*{Acknowledgment}
This work has been supported by NSERC Canada Research Chairs program, MITACS, and Ericsson.

\bibliographystyle{IEEEtran}
\bibliography{bibliography.bib}

\end{document}